%Paper: 9110021
%From: MANDAL%tifrvax.bitnet@pucc.PRINCETON.EDU
%Date: Mon, 7 Oct 91 21:59 IST

%%this texfile uses macropackage phyzzx
\input phyzzx
\hfill{ETH-TH-91/30}\break
\indent\hfill{IASSNS-HEP-91/52}\break
\indent\hfill{TIFR-TH-91/44}\break
%{
\indent\hfill{September, 1991}\break
\date{}
\titlepage
\title {GAUGE THEORY FORMULATION OF THE $C=1$ MATRIX MODEL :
SYMMETRIES AND DISCRETE STATES }
\author{Sumit R.~Das \foot{e-mail : das@tifrvax.bitnet},
Avinash Dhar \foot{e-mail : adhar@tifrvax.bitnet}}
\address {Tata Institute of Fundamental Research,
Homi Bhabha Road, Bombay
400 005, INDIA}
\author{Gautam Mandal \foot{e-mail: mandal@tifrvax.bitnet
(after 4 October, 1991).}$^{**}$}
\address{Institute for Advanced Study, Princeton, N.J. 08540, USA}
\author{Spenta R.Wadia \foot{e-mail :
wadia@iassns.bitnet}\footnote{**}{On leave from: Tata Insitute of
Fundamental Research, Bombay 400 005.}}
\address{Theoretische Physik, ETH-Honggerberg, 8093 Zurich, SWITZERLAND}
\centerline{and}
\address{Institute for Advanced Study, Princeton, N.J. 08540, USA}
\endpage
\abstract {We present a non-relativistic fermionic field theory in 2-dimensions
coupled to external gauge fields.
The singlet sector of the $c=1$ matrix model corresponds to a specific
external gauge field.
The gauge theory is one-dimensional
(time) and the space coordinate is treated as a group index. The
generators of the gauge algebra are polynomials in the single particle
momentum and position operators and they
form the group $W^{(+)}_{1+\infty}$. There are corresponding Ward identities
and
residual gauge transformations that leave the external gauge fields
invariant. We discuss the realization of these residual symmetries in
the Minkowski time theory and conclude that the symmetries generated
by the polynomial basis are not realized. We motivate and
present an analytic continuation
of the model which realises
the group of residual symmetries. We consider the classical
limit of this theory and make the correspondence with the discrete
states of the $c=1$ (Euclidean time) Liouville theory. We explain the
appearance of the $SL(2)$ structure in $W^{(+)}_{1+\infty}$. We also
present all the Euclidean classical solutions and the
classical action in the classical phase space. A possible relation of
this theory to the $N=2$ string theory and also self-dual Einstein gravity
in 4-dimensions is pointed out.}

%\centerline {(Submitted to )}

\endpage

\def\psdxz{\psi (x,0)}
\def\ps{\psi (x,t)}
\def\psd{\psi^{+}(x,t)}
\def\psdz{\psi^{+}(x,0)}
\def\half{{1 \over 2}}
\def\pat{\partial_t}

\def\pax{\partial_x}
\def\paxsq{\partial_x^2}
\def\inx{\int dx}
\def\inxt{\int dt~dx}
\def\e{\epsilon}
\def\cd{{\cal D}}

\def\pmd{d_{\pm}}
\def\Axyt{A_{xy}(t)}
\def\pst{\psi (t)}

\def\psxt{\psi_x (t)}
\def\psxdt{\psi_x^{+} (t)}
\def\psyt{\psi_y (t)}

\def\At{A(t)}
\def\cU{{\cal U}}
\def\bA{{\bar A}}

\def\PB{{\rm PB}}

\def\sqtwo{{\sqrt{2}}}

\def\cW{{\cal W}}

%%%%%%%%%%%%%%%%%%%%%%%
\def\zb{{\bar z}}
\def\del{\partial}
\def\poi#1#2{\{#1,#2\}_\PB}
\def\woneplus{W^{(+)}_{1+\infty}}

\Ref\BIPZ{E.Brezin, C.Itzykson, G.Parisi and J.Zuber, Comm. Math.
Phys. 59 (1978) 35.}
\Ref\BKZ{E.
Brezin, V. Kazakov and Al.B. Zamolodchikov, Nucl. Phys. B338 (1990)
673.}
\Ref\GROSSMILKOVIC{D. Gross and M. Miljkovic',
Phys. Lett. 238B (1990) 217.}
\Ref\GINZINN{P. Ginsparg, J.
Zinn-Justin, Phys. Lett. 240B (1990) 333.}
\Ref\PARISI{G. Parisi, Phys. Lett. 238B (1990)}.
\Ref\SW{A.M. Sengupta and S.R. Wadia,
Int. J. Mod. Phys. A6 (1991) 1961.}
\Ref\GKA{D. Gross and I. Klebanov,
Nucl. Phys. B352 (1990) 671.}
\Ref\MORE{G. Moore, Rutgers Preprint RU-91-12
(1991).}
\Ref\KS{D.Karabali and B.Sakita, City College Preprint,
CCNY-HEP-91/2.}
\Ref\KOST{I. Kostov, Phys. Lett. 215B (1988) 499.}
\Ref\SHAPIR{J. Shapiro, Nucl. Phys. B184 (1981) 218; A.Jevicki and B.Sakita,
Nucl. Phys. B165(1980) 511.}
\Ref\DJEV{S.R. Das and A.
Jevicki, Mod. Phys. Lett. A5 (1990) 1639}
\Ref\POLA{J. Polchinski,
Nucl. Phys. B346 (1990) 253}
\Ref\MSWA{G.Mandal, A. Sengupta and S.R. Wadia,
Mod. Phys. Lett. A6 (1991) 1465.}
\Ref\DJR{K. Demeterfi, A. Jevicki and J.P. Rodrigues, Brown Preprint
BROWN-HET-795 (1991).}
\Ref\GKB{D. Gross and I. Klebanov, Princeton University
Preprint PUPT-1242 (1991).}
\Ref\POLB{J. Polchinski, Texas Preprint,
UTTG-06-91 (1991).}
\Ref\DIF{P. Di Francesco and D.
Kutasov, Princeton Preprint PUPT-1237 (1991).}
\Ref\KAC{V.
V. Kac and D. Kazhdan, Adv. in Math. 34 (1979) 97; G. Segal, Comm. Math.
Phys. 80 (1981) 301; M. Wakimoto and H. Yamada, Hiroshima Math. J. 16
(1986) 427. }
\Ref\POLY{A.M. Polyakov, Mod. Phys. Lett. A6 (1991) 635.}
\Ref\GKN{D. Gross, I. Klebanov and M.J. Newmann, Nucl. Phys. B350
(1990) 621}
\Ref\DAG{U. Danielsson and D. Gross,
Princeton Preprint PUPT-1258 (1991).}
\Ref\MSWB{ G. Mandal, A. Sengupta and S.R. Wadia, Mod. Phys. Lett. A6
(1991) 1685.}
\Ref\WITA{E. Witten, Phys. Rev. D44 (1991) 314.}
\Ref\ROC{M.Rocek, K.Schoutens, and A.Sevrin, IAS Preprint,
IASSNS-HEP-91/14}
\Ref\ELITZUR{S.Elitzur, A.Forge and E.Rabinovici, Hebrew Univ.
Preprint RI-143/90.}
\Ref\DDMW{S.R. Das, A. Dhar, G.
Mandal and S.R. Wadia (unpublished).}
\Ref\AVJEV{J. Avan and A.
Jevicki, Brown Preprint BROWN-HET-801 (1991).}
\Ref\POLC{J. Polchinski,
Texas Preprint UTTG-16-91 (1991).}
\Ref\POPE{C.N.Pope, L.J.Romans and X.Shen, ``A brief History of
$W_\infty$,'' in {\sl Strings 90}, ed. R.Arnowitt et al (World
Scientific, 1991).}
\Ref\BALA{A.P. Balachandran, G. Marmo, B.S. Skagerstam and A. Stern,
"Gauge Symmetries and Fiber Bundles", Springer Lecture notes in Physics
vol. 188}
\Ref\VAFA{H. Ooguri and C. Vafa, Univ. of Chicago and Harvard
PreprintEFI-91/05, HUTP-91/A003 (January, 1991).}
\Ref\MORENEW{G.Moore and N.Seiberg, Rutgers Preprint.}
\Ref\AVJEVNEW{J.Avan and A.Jevicki, Brown Preprint, Brown-HET-824
(1991).}
\Ref\WITB{E. Witten, IAS Preprint, IASSNS-HEP-91/51.}
\Ref\RAJV{S. Rajeev, Phys. Lett. B209 (1988) 53; Phys. Rev. D42 (1990) 2779.}

\chapter{ Introduction}

In this paper we explore the $c=1$ matrix model with the optimism that
such a study may hint at some general symmetry principles of string theory.

Let us begin by recalling some of the known results of the $c=1$
matrix model and the $d=2$ critical string theory relevant to this
paper. In the double scaling limit the matrix model in the
$U(N)$-invariant sector is equivalent to the theory of free fermions
in one space and one time dimensions interacting with a background
potential $V(x) = - \half x^2$ [\BIPZ-
\PARISI]. This system has a natural  and exact description in terms of a
two-dimensional non-relativistic
field theory of fermions [\SW,\GKA,\MORE,\KS]. The elementary
low-energy
excitation of this model is a particle-hole pair which has a massless
dispersion relation [\KOST,\SHAPIR] and has been identified with the
tachyon of the continuum theory [\DJEV,\SW].
In the continuum theory, the physical spectrum corresponds to
exciations around a specific tachyon background [\POLA].
Scattering amplitudes of these tachyons have been calculated by
several authors [\MSWA,\DJR,\MORE,\GKB,\POLB].
These results are in agreement with the
calculation of the scattering amplitude in two dimensional critical
string theory [\DIF] (where the liouville mode acts as a spatial
dimension) with a flat metric and a linear dilaton.

Besides the massless tachyon the spectrum of the two dimensional
critical string theory (in the above mentioned background) has an
infinite tower of discrete states labelled by two non-negative
integers $(r,s)$ [\KAC-\GKN] These states are the gauge invariant
content of the higher spin states. In higher dimensions one is left
with {\it fields} after all the gauge degrees of freedom are removed
from the higher spin gauge fields. However, in two dimensions one is
left with a discrete set of states. The existence of such states in
the matrix model has been most clearly indicated by
a study of the two point function of time dependent operators [\DAG].

In what follows we study the one dimensional matrix model in an
enlarged framework which enables us to organize the above facts.
Hopefully this framework will throw light on the understanding of the
two dimensional black hole [\MSWB,\WITA,\ROC,\ELITZUR]
in the matrix model
and other important issues in string theory such as symmetry
principles and a background independent formulation.

Let us summarise our results.

In section 2, we review the infinite number of conservation laws [\SW,
\GKA] in the fermion field theory and write down the conserved currents.
We show that the conservation laws are a reflection of global
symmetries in which the fermion field not only gets multiplied by some
function of $x$ but at the same time gets acted on by the translation
operators involving powers of $-i\del/\del x$.  This observation
motivates us to consider (section 3) a fermion field theory coupled to
external gauge fields such that the theory has a symmetry group which
include arbitrary combinations of translation and phase multiplication
on the fermi field (to be precise, $\delta\psi(x,t) \to
\sum_{n=0}^{\infty} \e_{n}(x,t)p^n\psi=\sum_{m,n=0}^{\infty}
\e_{mn}(t) x^m p^n\psi(x,t)$).
Interestingly the above set of gauge transformations are precisely the
set of all unitary transformations in the quantum mechanical Hilbert
space of a single particle! Clearly such transformations form a group
(in section 5 we identify the Lie algebra to be $\woneplus$).  We
write down the gauge-invariant lagrangian in the quantum mechanical
notation. We show that the $c=1$ double scaled matrix model
corresponds to a particular choice of the background gauge field. In
section 4 we derive the Ward identities of gauge invariance. In
section 5 we consider a fixed arbitrary background and find the
special gauge transformations that leave the background gauge field
invariant.  These do not lead to Ward identities but rather act as
symmetry transformations. Their generators are constants of motion for
an arbitrary time dependent background and as expected satisfy the
$\woneplus$ algebra.  We specifically study the background gauge field
$\bar A=-1/2 (p^2- x^2)$, which corresponds to the martix model and
present the specific time-dependence of the special
gauge transformations.

In section 6, we discuss the algebra of the above symmetry
transformations in the fermion Fock space. We find that in the
Minkowski theory where the single particle wave functions are
parabolic cylinder functions, the unitary group as generated by the
differential operators $x^mp^n$ does not have a well defined action on
the Hilbert space, meaning that each element of this basis
transforms the single particle
wavefunctions out of the Hilbert space.
In sections  7 and 8,  we
motivate and present  a specific analytic continuation of the
theory, viz. $t\to it$ and $p\to -ip$ (and $A(t)\to -i A(t)$)
which enables the construction
of a representation of the symmetry algebra generated by the basis $x^mp^n$.
The background gauge
field in this case is $\bar A= 1/2(p^2+ x^2)$, the ordinary harmonic
oscillator hamiltonian. We show that all the states of the fermion
field theory can be generated by acting with the symmetry generators
on the Fermi sea. We emphasize that the
analytically continued fermion theory does not contain any state with
continuous energy: even the tachyon spectrum is discrete.

In section 9 we
discuss the classical limit as $g_{st}\to 0$ using the
fermi fluid picture
recently emphasized by Polchinski[\BIPZ, \POLB].
We  show that the classical limit of
the expectation value of the generator $W^{(r,s)}$ in a given state $|f>$,
described by a profile $f(x,p)=0$ of the fermi fluid, can be expressed
as the phase space average of a function $V_{rs}(\tau,t-\theta)$:
$$ V_{rs}(\tau,\theta)= \exp(-2\tau) \exp(-\tau(r+s)+ i\theta(r-s))$$
where $x=\exp(-\tau)\cos\theta,\; p=\exp(-\tau)\sin\theta$
parametrise  the single-fermion phase space. We also discuss how
$SL(2)$ naturally arises in the discussion of the single-particle
phase space, and present a simple understanding of the representation
theory in classical terms. Finally, we present a complete list of
the Euclidean
classical solutions of the theory and propose this space of solutions
as the classical phase space. We identify the classical limit of the
field theory operators $W_{rs}$ as functions  $w_{rs}$ on this space.
We calculate the Poisson brackets of these functions; we show that all
other functions on the phase space are functions of these basic
observables $w_{rs}$ and hence the knowledge of $\poi{w_{rs}}{w_{mn}}$
consitutes a complete specification of the symplectic form on this
phase space. Since the hamiltonian in this phase space is also known
(it is simply $w_{11}$) we have a complete specification of the
classical physics. We present the classical action in this phase space.

In section 10, we consider   perturbing the model by the operators
$\sum_{rs} g_{rs}W^{(r,s)}$ which is a gauge-inequivalent deformation
of the background gauge field. We briefly discuss a scenario for a
background independent formulation of the theory.

In section 11, we point out the possible connection of this work with
that of Vafa and Ooguri [\VAFA] on the $N=2$ string field theory and
also self-dual Einstein gravity in 4-dimensions.

While this work was in progress we received
several papers which partly overlap with this work
[\MORENEW,\AVJEVNEW, \WITB].
The gauge group discussed in this paper
has also been independently discussed by S. Rajeev [\RAJV].

\chapter{Fermion field theory}

The mapping of the one dimensional matrix model onto a theory of
non-relativistic fermions in a potential is a most fortunate yet
mysterious circumstance, especially when viewed from the viewpoint of
the two dimensional string theory. The double scaled fermion field
theory is described by the action $$ S = \inxt \psd [ i \pat + \half
\paxsq + \half x^2] \ps \eqn\one$$
Since the fermions are
non-interacting the energy of each fermion is separately conserved.
This implies that the sum over any power of the individual energies is
also conserved. These conserved charges are given by $Q_n = \inx \psd
h^n \ps,~~~n = 0,1,2 \cdots$ where $h = \half(p^2 - x^2)$ and $p = - i
\pax$ [\SW, \GKA].  It is easy to find local conservation laws with
charge and current densities given by [\DDMW,
\AVJEV,\POLC]
$$ J_n^0 = \psd h^n \ps ,~~~~~J_n^1 = {i \over 2} ( \pax\psd h^n \ps -
\psd \pax h^n \ps) \eqn\two$$ which satisfy $\pat J_n^0 - \pax J_n^1 =
0$ by the equations of motion.  These conservation laws are statements
of the global symmteries of the fermion action \one $$ \delta \ps = i
\sum_{n=0}^{\infty} \alpha_n h^n \ps~~~~~~
\delta \psd = -i \sum_{n=0}^{\infty} \alpha_n h^n \psd \eqn\four$$

For reasons which will be clear later, we will now enlarge the
framework of our discussion by considering a model in which these
symmetries are {\it gauged}. This means we introduce gauge fields
which couple to these currents and assign transformation rules to them
so that the transformations in \four\ are symmetries with parameters
$\alpha_n(x,t)$ which are arbitary functions of $x$ and $t$.
The transformations in \four, when
$\alpha_n$'s are constants, involve {\it specific}
linear combinations of operators of the form $x^m \pax^n$ acting on
the fermion fields. We shall consider instead a model in which the
gauge transformations involve {\it arbitrary} combinations of these
operators.

\chapter{Gauge theory of the group of unitary transformations in a Hilbert
speace}

The framework hinted at in the previous section may be best described in
the following way. The fermion field $\ps$ is viewed as a vector in a
Hilbert space ${\cal H}$ such that
$$ \ps \equiv \psi_x (t) = <x | \psi (t)> $$
The index $x$ labels the component of the vector. In the following we shall
sometimes denote $|\psi(t)>$ by $\psi(t)$. Now consider the action of
unitary operators ${\cal U}$ on $\psi$, $\psi \rightarrow {\cal U} \psi$.
This is clearly a symmtery of the free fermion theory
$$ S_0 = \int dt < \pst | i \pat | \pst > = \inxt ~\psxdt i
\pat \psxt \eqn\six$$
The symmetry may be gauged by introducing a self-adjoint gauge field $\At$
($\At = \At^{+}$). The action
$$S = \int dt <\pst|i \pat + \At  | \pst> \eqn\seven$$
is then gauge invariant under the transformations
$$ \eqalign{&|\pst> \rightarrow {\cal U}(t) |\pst> \cr &
\At \rightarrow \cU (t) \At \cU^{+}(t) + i \cU (t) \pat \cU^{+} (t)
}\eqn\eight $$
Clearly the set of unitary transfromations form a group. $\cU$ may be
parameterized as
$\cU = {\rm exp} (i\e)$ where $\e =  \e^{+}$.

We can realize the hilbert space ${\cal H}$ in terms of the space of
functions on the real line $R^1$. Then $A(t)$ can be considered as a
function of the basic operators $x$ and $p$ which satisfy the
commutation relations $[x,p] = i \hbar$. The components of $\At$ are obtained
by expanding it in terms of the set of self adjoint operators $(x^mp^n +
p^nx^m);~ m,n \ge 0$.
$$A(t;x,p) = \sum_{m,n = 0}^{\infty} A_{mn}(t) (x^mp^n + p^nx^m) \eqn\nine$$
The operators ${\hat l}_{mn}= x^m p^n$ form a closed algebra:
$$[{\hat l}_{mn},{\hat l}_{rs}]= \sum_p(-i)^p (C(n,p)r_p-C(s,p)m_p){\hat
l}_{m+r-p,n+s-p}  \eqn\ninea$$
Here $n_p\equiv n(n-1)\cdots (n-p+1), C(n,p)=n_p/p!$ when $n\ge p$,
and $0$ otherwise. The sum over $p$ is through non-negative integers,
there are only finite number of terms because the coefficients vanish
for large enough $p$ by the above definitions.
This  algebra is actually isomorphic to the algebra called $\woneplus$
(see section 5).
If we choose the representation $p = - i \pax$ then
$$\Axyt = <x| A(t;x,p)|y> = \sum_{m,n=0}^{\infty} A_{mn} (t) [(x^m  +
y^m) (i\pax)^n] \delta(x-y) \eqn\ten$$
A similar expansion holds for the parameter $\e$ of gauge transformations,
where $\cU \sim 1 + i\e$ and $\e$ is hermitian. One has
$$ \e_{xy} = <x| \e (t) |y> = \sum_{n=0}^{\infty}
[(\e_n(x)  + \e_n(y)) (i\pax)^n] \delta (x-y) \eqn\eleven$$
We also define $\e_{mn}$ by
$$ \e_n(x) = \sum_{m=0}^{\infty} \e_{mn}x^m $$
The infinitesimal gauge transformation \eight\ may be written in component
form as
$$\delta \ps =  i\int dy \e_{xy}(t) \psi (y,t) ~~~~
\delta \psd =  -i\int dy  \psi^{+} (y,t) \e_{yx}(t)\eqn\twelve$$
and
$$\delta \Axyt = i\cd \e = i(i \pat \e (t) + [ A(t) , \e(t)])_{xy}
\eqn\thirteen$$
In terms of these components the action becomes
$$ S = \inxt~ dy~ \psxdt [ i \pat \delta(x-y) + \Axyt ] \psyt \eqn\fourteen$$

The fermionic field theory description of the singlet sector of the $d=1$
matrix model is related to this gauge theory in the following way. The
fermion field theory \one\ may be easily seen to be the gauge theory
\fourteen\ in a {\it specific} and {\it fixed} background gauge field
given by $\bA$
$$ \bA (t;x,p) = - h = -\half (p^2 - x^2) \eqn\fifteena$$
or, in terms of components
$$ \bA_{xy}(t) = <x|A(t;x,p)|y> = \half (\pax^2 + x^2) \delta(x-y)
\eqn\fifteenb$$
Thus we shall be concerned with the above formulated gauge theory in {\it
fixed} backgrounds.

In the classical limit as $\hbar \rightarrow 0$ ($\hbar$ plays the role of
string
coupling) we have the usual correpsondence. The commutator $[x,p] = i\hbar$ is
replaced by the Poisson bracket and $x$ and $p$ are coordinates on a phase
space. Operators like the gauge fields then become functions on the
classical phase space and the algebra of unitary transformations goes over
to the algebra of area preserving diffeomorphisms of the phase space. This
algebra is generated via Poisson brackets, the generators $l_{mn} = x^n
p^m,~~n,m \ge 0$ satisfy
$$ \lbrace l_{mn}, l_{rs} \rbrace_{\rm PB} = (ns-rm) l_{n+r-1, m+s-1}
\eqn\sixteenc$$

\chapter{Ward Identities}

We now turn to some consequences of the gauge symmetries, viz. Ward
identities. Consider the functional integral:
$$ {\cal Z} [\bA] = \int \cd \psxdt \cd \psxt e^{{i \over h} \int dt
<\pst| i \pat + \bA |\pst>} \eqn\seventeen$$
where $\bA$ is an arbitrary background gauge field.
Ward identities
are a consequence of the invariance of the fermion measure under gauge
transformations. We regard the fermion measure to be invariant because we
do not expect any anomalies for non-relativistic fermions. Since the
action is gauge invariant, we have the identity
$$ {\cal Z} [\bA] = {\cal Z} [ \bA + i{\bar \cd} \e] \eqn\eighteen$$
where the covariant derivative $\cd$ has been defined in \thirteen\ and
the bar means that the gauge field in $\cd$ is the background gauge field
$\bA$.

Introducing the notation
$$ R_{yx} (t) = {\delta \over \delta \Axyt} {\cal Z} \eqn\eighteena$$
\eighteen\ can be written as a differential equation
$$ \inxt~dy~ {\bar \cd}\e_{xy} (t) {\bar R}_{yx}(t) = 0 \eqn\nineteen$$
For gauge transformations which {\it do not} keep the background gauge
field invaraint, i.e. ${\bar \cd} \e \neq 0$, we can integrate by parts in
\nineteen\ and arrive at the Ward identities or "transversality conditions"
$$ i \pat {\bar R}_{yx} + [ \bA (t), {\bar R} (t)]_{yx} = 0
\eqn\twenty $$

In the special case when $\bA=-(p^2-x^2)/2$ the fermion theory
corresponds to the standard double scaled matrix model.
These identities then
imply an infinite set of relations between correlation
functions of that model.
In this paper we shall not pursue further equation \twenty\
especially with regard  to important questions of boundary conditions and
the ability to explicitly evaluate the correlation functions.

\chapter{ Residual Gauge Symmetry: $\woneplus$}

Given a particular background gauge field $\bA(t)$, a generic gauge
transformation $\epsilon(t)$ would not leave the gauge field
invariant. It is an interesting question to ask what is the set of
gauge transformations $\e(t)$ that do leave the background invariant.
Clearly such gauge transformations would form a subgroup of the
original gauge transformations, for instance in case of the
'tHooft-Polyakov monopole the residual subgroup was a $U(1)$
subgroup of the original gauge group $SO(3)$.

These special gauge transformations satisfy
$${\bar \cd} \e = i \pat \e + [ \bA(t), \e] = 0 \eqn\newfiveone$$

The general solution of \newfiveone\ is given by
$$\e(t)=U(t)\e(0) U(t)^{-1} \eqn\newfivetwo $$
where $U(t)= T\exp i\int ^t \bA(t') dt'$ is a unitary operator and
$\e(0) \equiv \epsilon (x,p)$ is as yet an arbitrary operator on the
single particle Hilbert space.

The generators of the special gauge transformations in the field theory are
given by
$$W[\e,t]=\int dx \psd \e_{xy}(t) \psi(y,t)= <\psi(t)|\e(t)|\psi(t)> \eqn
\newfivethree $$
Using \newfiveone\ and the equation of motion of $\ps$, it is easy to
check the expected result $dW[\e,t]/dt=0$.

We now make the specific choice of
expanding it in terms of the basis of generators ${\hat l}_{mn}= x^m
p^n$. This basis is well defined and general enough. In particular it
includes the single particle Hamiltonian and the generators of the Lie
algebra $SL(2)$.
In terms of this basis we have $W(\e(t))= \sum_{mn} \tilde
\e_{mn}  W_{mn}(t)$ where
$$W_{mn}(t)= \int dx \psdz {\hat l}_{mn}(t)\psdxz \eqn\newfivefour$$
and we have defined
${\hat l}_{mn}(t)= U_t {\hat l}_{mn} U_t^{-1}$. The constants of
motion satisfy the same algebra as that of the single-particle
generators ${\hat l}_{mn}$. Therefore the residual symmetry algebra
around any background is the same.

We now discuss the case of the background \fifteena\ which is
explicitly time-independent. In that case $U_t= \exp(i \bA t)$ and we
can look for solutions of the form $\e(t)= \exp (iEt)~ \e(0)$.

Equation \newfiveone\ then becomes the eigenvalue problem
$$ {\rm ad}~\bA \cdot \e = E ~ \e \eqn\twentytwo$$
where ${\rm ad}~\bA$ denotes the adjoint action of $\bA$. It is convenient
to introduce the eigenoperators $\pmd \equiv {1 \over \sqtwo}
(x \pm p)$ such that
$${\rm ad}~\bA \cdot \pmd = \pm i \pmd \eqn\twentythree$$
Now since the action of ${\rm ad}~\bA$ is associative, we have
$${\rm ad}~\bA \cdot (\pmd)^m = \pm i~m~(\pmd)^m \eqn\twentyfour$$
where $m$ is a non-negative integer. A solution of \newfiveone\ is,
therefore, labelled by two positive integers and can be written as
$$ \e^{rs}_{xy} (t) \equiv \e^{rs}(t;x,y) = \half \lbrace E_+^r,E_-^s
\rbrace_x \delta (x-y) \eqn\addone$$
where we have defined
$$ E_{\pm} \equiv e^{\mp t} d_{\pm} \eqn\addtwo$$

The hermitian charges that generate the special gauge transformations
are given by \newfivethree\ as
$$W^{(r,s)} = \inx \psxdt \e_{rs}(t;x,y) \psyt
\eqn\twentyninea$$
They can be expressed as a linear combination of the $W_{mn}$ of
equation \newfivefour.

Let us now make correspondence with similar algebras that exist in the
literature.
The operators $W^{(r,s)}$ form a closed algebra under commutation, which
is identical to the algebra of the single particle operators $\e_{rs}
(t)$. This is the algebra $\woneplus$. The generators of the
standard $\woneplus$ algebra are linear combinations of the
$W^{(r,s)}$ . These can be constructed following the construction of $W_{1
+ \infty}$ algebra as an enveloping algebra of a $U(1)$ Kac-Moody algebra
with a derivation given in [\POPE].
In the present case we take the modes
of the $U(1)$ current to be $E_{-}^m,~~m \ge 0 $ and for the derivation we
take i times the single particle hamiltonian, $ih = {i \over 2}(p^2 -
x^2) = - {i \over 2} \lbrace E_{+},E_{-} \rbrace$. These satsify the
requirements of the construction given in [\POPE].
Then using the notation
$V_m^j$ for the $W_{1 + \infty}$ generators where the spin is $(j+2)$ and
the mode is $m$ we have the recursion relation whcih determines them
$$ V_m^j = (ih + {m \over 2})V_m^{j-1} + { j^2 (j^2 - m^2) \over 4(4j^2 -
1)} V_m^{j-2} \eqn\thirty$$
where
$$V_m^{-1} = E_{-}^m \eqn\thirtya$$
and
$$V_m^0 = (ih + {m \over 2})E_{-}^m = - {i \over 2} \lbrace E_{+},
E_{-}^{m+1} \rbrace \eqn\thirtyb$$
For example
$$V_m^1 = - \half \lbrace E_{+}^2, E_{-}^{m+2} \rbrace - {1 \over 3} (m+1)
(m+2) E_{-}^m \eqn\thirtyc$$
It is clear that in this way we can express any $V_m^j$ as a linear
combination of the $\e_{rs}$. Since $r$ and $s$ are restricted to be
non-negative we see that the $V_m^j$ thus obtained are restricted to $ j
\ge -1$ and $m \ge -j +1$. In this way we can construct the standard
$\woneplus$ algebra from linear combinations of $W^{(r,s)}$.

We mention some other important properties of the operators $W^{(r,s)}$.

\noindent (i) $W^{(r,s)}$ are time independent. It turns out the these
conservation laws are implied by the local conservation of the following
currents
$$\eqalign{ &J^0_{r,s} (x,t) = \psd  \e_{rs} \ps  \cr &
J^1_{r,s} (x,t) = { i \over 2} [ \pax \psd \e_{rs} \ps - \psd \pax \e_{rs}
\ps ] \cr &
\pat J^0_{rs} - \pax J^1_{r,s} = 0} \eqn\thirtytwo$$

\noindent (ii) For $r=s,~~~J^{\mu}_{r,r}$  can be expressed as linear
combinations of the currents given
in \two. The corresponding charges are the set $Q_r = W^{(r,r)}$. Clearly
these charges commute among themselves
$$ [ W^{(r,r)},W^{(r',r')}] = 0 \eqn\thirtythree$$
and $-W^{(1,1)}$ is the hamiltonian.

\noindent (iii) \underbar{The $SL(2)$ in }$\woneplus$:

The operators $W^{(r,s)}$ have an interesting $SL(2)$ structure. The
$SL(2)$ structure has appeared in the discussions of the continuum
theory in [\DAG]. In
fact, it can be shown that linear combinations of the
$W^{(r,s)}$ fall into
$SL(2)$ multiplets. The $SL(2)$ is generated by $$ J_{+} = {i \over 2}
W^{(2,0)},~~~~J_{-} = {i \over 2}W^{(0,2)},~~~ J_3 = - {i \over 2}
W^{(1,1)} \eqn\fiveone$$ They satisfy the standard algebra $$
[J_+,J_-] = -2 J_3,~~~~~[J_3,J_{\pm}] = \pm J_{\pm} \eqn\fivetwo$$ and
the quadratic casimir is $$ C_2 = J_3^2 - \half(J_+ J_- + J_- J_+)
\eqn\fivethree$$ The set of operators that form an $(n+1)$ dimensional
representation of the $SL(2)$ may be constructed as follows. One
starts with the operator $W^{(n,0)}$. The next member of the set is
given by the commutator $[J_-,W^{(n,0)}]$. Taking a further commutator
gives the next member $[J_-,[J_-,W^{(n,0)}]]$ and so on. One can
easily show that the last operator in the chain is $W^{(0,n)}$ and so
the chain stops after $n$ steps. The resulting set of $(n+1)$
operators forms the $(n+1)$ dimensional representation of $SL(2)$ by
construction. (One could have equivalently started with $W^{(0,n)}$
and obtained the above multiplet by the repeated action of $J_{+}$).

Let us denote a member of a given multiplet by $\cW^{(r,s)}$. This is
obviously a linear combination of the $W^{(r,s)}$. For example,
$$ \eqalign{ \cW^{(n,0)} = & W^{(n,0)},~~~\cW^{(n,1)} = W^{(n,1)}, \cr &
\cW^{(n,2)} = W^{(n,2)} - 2n(n-1) W^{(n-2,0)} \cdots }\eqn\fivefour$$
It may be easily checked that the $J_3$ eigenvalue of $\cW^{(r,s)}$ is
$({r-s \over 2})$ while the quadratic casimir on it gives the eigenvalue
$({r + s \over 2})({r+s \over 2} + 1)$. The operators $J_3$ and the
quadratic Casimir act by adjoint action on $\cW^{(r,s)}$. Thus the two numbers
 that specify
any member $\cW^{(r,s)}$ of a given $SL(2)$ multiplet are $(r-s)$ and
$(r+s)$ respectively.

\chapter{On  representation of $\woneplus$
in the fermion field theory in an inverted harmonic oscillator potential}

In the previous section we considered the algebraic structure of the
group of residual gauge transformations. Now we consider the important
question of its representation in the fermionic field theory.

As is well-known, the fermionic field theory \one\ can be built
entirely from the knowledge of the single-particle states. These are
given by parabolic cylinder functions [\BKZ,\MORE].
The ground state $|\Omega>$ of
the  field theory is obtained by filling all the single-particle
levels upto the fermi level $\mu$. Let us call the distribution of
single-particle energy levels $\rho(E)$. Using this we can easily calculate the
 energy of the ground state; in fact we
can compute the eigenvalues of all the commuting generators
$W^{(r,r)}$ which are linear combinations of the form $ \sum_{l=0}^r c^r_l
\int\psd h^l \ps$, where $c^r_l$ are constants,
in the ground state by computing moments of
$\rho(E)$. The results are, after subtracting an infinite constant:
$$e_r=\sum_{l=0}^r c^r_l \int_\mu^{\infty} dE\rho(E) E^l$$

Now consider the action of the generator $W^{(r,s)}, r\not=s$ on the
ground state:
$$|r,s>= W^{(r,s)}|\Omega> \eqn\sixone$$
Using the commutation relation $[W^{(1,1)}, W^{(r,s)}]= i(r-s)
W^{(r,s)} $, which is basically a refelction of the single-particle
relation \twentythree, we see that
$$W^{(1,1)}|r,s>= (e_1+ i(r-s))|r,s> \eqn\sixtwo$$
Hence  the hamiltonian of the theory, which is just $-W^{(1,1)}$, has a
complex eigenvalue in the state $|r,s>$. Since on the other hand we
can explicitly construct all the states of the field theory in terms
of multiple electron-hole excitations and they all have real energies,
(remember  parabolic cylinder functions have real energy eigenvalues),
this means that the state $|r,s>$ cannot be expressed as a linear
combination of the complete set of states in the field theory; in
other words the $\woneplus$ transformation of the ground
state takes it out of the Hilbert space.

Let us explain the last comment in a little more detail. Let us expand the
 second quantised fermi field in terms of parabolic
cylinder wave-functions and the corresponding creation-annihilation
operators of the single-particle states. Now the action of $W^{(r,s)}$
on the ground state would generically involve applying the operators
$d_{\pm}= {1 \over \sqtwo} \left(x \mp i\del/\del x \right)$
on parabolic cylinder functions. If one
looks up the asymptotic behaviour of parabolic cylinder functions with
energy eigenvalue $\lambda\in R$,
one finds that they behave like $\psi\sim 1/\sqrt{x} \exp if(x)$ where
$f(x)= {\rm constant}\; x^2+{\rm constant}\; \lambda log(x)+\cdots$
The oscillatory behaviour is characteristic of the fact that these
represent scattering states.
It is easy to see that the action of the operators $d_\pm$ amounts to
changing the eigenvalue $\lambda$ by $\pm i $ which implies that even
though $\psi$ vanishes at $\pm \infty$ the boundary conditions of the
new eigenfunctions obtained this way are different. For example the action
of $d_+$ results in a new wavefunction which blows up at $\infty$. Clearly
such a wave-function cannot be expressed as a linear combination of
$\psi(\lambda, x)$'s for real $\lambda$'s. We say that $d_+$ does not
have a well defined action on the Hilbert space of the parabolic
cylinder functions. The same is true for a generic differential
operator formed as a finite linear combination of $d_+^r d_-^s$ or
equivalently $x^mp^n$.

At this point let us recall the discussion after equation \newfivetwo. There
the operator $\epsilon (x,p)$ was unspecified and we made a choice of
expanding it in terms of ${\hat l}_{mn}$. Notwithstanding the virtues
of this basis one can certainly imagine a class of operators $\epsilon
(x,p)$, closed under commutation, which have well defined action in
the Hilbert space of parabolic cylinder functions. An example are the
operators $f_{\alpha\beta}(x,p) = \exp (i\alpha x + i \beta p)$ for
$\alpha$ and $\beta$ arbitrary real numbers. From the viewpoint of the
classical limit (which will be subsequently discussed in section 9)
one is distinguishing between generators of canonical transformations
which are expressed as linear combinations of the sets $F = \{
f_{\alpha\beta}(x,p) = \exp (i\alpha x + i \beta p)\}$ and $L = \{
{\hat l}_{mn}(x,p) = x^mp^n\}$.

In the next section we shall see that the basis L indeed has a well
defined action in the Hilbert space of Hermite functions, basically
because these functions represent bound states and have an exponential
decay at infinity.

\chapter{Analytic continuation of the fermion field
theory}

In this section we motivate and discuss the analytic continuation of
the fermion field theory in which the algebra of residual gauge
transformations can be realised in the $L$ basis.

Let us recall that the fermion field theory \one\ can be expressed as
a Feynman path integral over the classical trajectories of the
fermions, which are governed by the action
$$iS_M= i\int dt ({\dot x^2\over 2}+ {x^2\over 2})
\eqn\sevenone$$
The classical equation of motion $d^2 x/dt^2-x=0$ is solved by the
hyperbolic functions $x(t)= A \cosh(t+\theta)$. The canonical momentum
is given by $p(t)=\dot x(t)=A\sinh(t+\theta)$. The hamiltonian is then
defined by $h(x,p)=p\dot x-(\dot x^2+ x^2)/2$, and evaluates to
$h(x,p)= (p^2-x^2)/2=-A^2/2$. Hence constant energy trajectories in
phase space are given by hyperbolas.

Now consider the analytic continuation of time $t\to it$ to a
Euclidean picture. The Euclidean action corresponding to the Minkowski
action \sevenone\ is
$$S_E=\int dt(-{\dot x^2\over 2}+{x^2\over 2})\eqn\seventwo$$
The classical equation of motion $\partial_t^2x +x=0$ (simple harmonic
oscillator) are solved by periodic functions $x(t)=A\cos(t+\theta)$.
The canonical momentum is given by $p=\del h/\del {\dot x}=-\dot x=
A\sin(t+\theta)$ and the hamiltonian is $h(x,p)= p\dot x+ \half(\dot x^2 -
x^2) = -1/2 (p^2+x^2)= -1/2 A^2$. The constant energy trajectories are
given by circles of radius $A$.

{}From the above discussion we deduce that the standard analytic
continuation of time $t\to it$ in the coordinate space formulation
corresponds to the analytic continuation $t\to it$ and $p\to -ip$ in
the phase space formulation. The analytic continuation is illustrated
in Figs 1a and 1b.

In the quantum theory the orbits are appropriately quantised and the
ground state is obtained by filling the fermi sea. In figures 1a and
1b we have indicated this by the shaded regions $x^2-p^2\le \mu$ and
$x^2+p^2 \geq\mu$ respectively. $\mu$ is the fermi energy and defines
the string coupling $g_{st}=1/\mu$. In the classical limit $g_{st}\to
0$ the states of the field theory can be described in terms of a fermi
fluid and we shall investigate in detail the classical solutions
(instantons) of the Euclidean field theory in terms of motion of this
fluid in section 9.

We wish to emphasize that in quantising the classical phase space (Fig
1b) obtained by the analytic continuation $t\to it$ and $p\to -ip$, we
are going beyond the standard Euclidean continuation of quantum
mechanics. The reason for this is that in the standard Euclidean
continuation, the quantum hamiltonian, obtained by the transfer matrix
method, does not change. The analytic continuation we have performed
changes the quantum hamiltonian from that of the inverted harmonic
harmonic oscillator $h=1/2(p^2-x^2)$ to that of the ordinary harmonic
oscillator $h=-1/2(p^2+ x^2)$. The final result of our analytic
continuation is identical (upto the overall sign of the hamiltonian)
to the result obtained by Gross and Milkovic [\GROSSMILKOVIC]
who regarded the
inverted harmonic oscillator as the usual hamonic oscillator with an
imaginary frequency. Tne verity of their procedure and also ours is
well supported by the fact the correlation function calculation in
Danielssen-Gross [\DAG] agrees completely with that of Moore [\MORE].

We end this section by discussing the single-particle levels and the
fermi sea. The single-particle levels are the hermite functions
$H_n(x)= <x|n>$ and the energy levels are $E_n=-(n+1/2), n=0,1,2,
\cdots $. The fermi sea is filled up from $n=\infty(E_n=-\infty)$ to
some level $n=n_F (E_n\equiv E_F=-\mu)$. The semiclassical
limit as $\mu\to\infty$ was described before (fig 1b). It is worth
commenting that in Fig 1b the number of unoccupied levels is also of
order $\mu$. This should be contrasted with the fact that the number
of unoccupied levels in Fig 1a (Minkowski picture) is actually
infinite. Hence the analytic continuation seemingly reduces the number
of degrees of freedom. This seems to be the analogue for functional
integrals of a similar phenomenon that occurs in the evaluation of
real integrals as a sum over poles.

\chapter{Representation of $\woneplus$ in the
analytically continued fermion field theory}

Most of the algebraic steps of section 5 in the single-particle
quantum mechanics can be repeated with the substitution $p\to -ip$. In
particular $d_\pm={1 \over \sqtwo}(x\pm p)\to {-i \over \sqtwo}(ix\pm p)$. Let
us denote $p-ix=a$ and
$p+ix=a^\dagger$ in the standard way. Since $t \to it$, the gauge
field $A$ is also analytically continued $A\to -iA$ so that $ dt\; A$
is left unchanged. Hence the residual gauge transformation (similarly
to \newfiveone) satisfies
$$i\del_t\e+[\bar{A},\e]=0 \eqn\eightone $$

Subsituting $\bar{A}=(p^2+x^2)/2=-h$ in \eightone\ we get
$$i\del_t\e- [h,\e]=0 \eqn\eighttwo$$

Once again, we write $\e(t)=\exp(iEt)\e(0)$ and get the eigenvalue
problem
$$-{\rm ad} h.\e=E\e \eqn\eightthree$$
Now since $(p^2+x^2)/2= (a^\dagger a+ a a^\dagger)/4$ therefore
$-{\rm ad}h.a=a$ and $-{\rm ad}h.a^\dagger=- a^\dagger$. Hence

$$ -{\rm ad}h.a^r a^{\dagger s}= (r-s)a^r a^{\dagger s}
\eqn\eightfour$$
Denote $a^ra^{\dagger s}=\e_{rs}$ and note that
$\e_{rs}=\e^\dagger_{sr}$. The general solution of \eighttwo\ is now
given by
$$\e(t)= \sum_{r,s} e^{i(r-s)t} \e_{r,s}\alpha_{r,s}\eqn\eightfive$$
where $\alpha_{r,s}^*=\alpha_{s,r}$ ensuring that $\e^\dagger= \e$.

The generators of the residual gauge transformation in the fermi field
theory are given by
$$W^{(r,s)}=\int dx \psd \e_{rs} \ps,\; W^{(r,s)\dagger}=W^{(s,r)}
\eqn\eightsix $$
They are constants of motion. The parameters $\e_{rs}=a^r
a^{\dagger s}$ again satisfy the $\woneplus$
algebra and imply the same
for $W^{(r,s)}$.

We can realise  $W^{(r,s)}$ in terms of
the constituent modes of the fermion field which can now be expanded in
terms of
hermite polynomials $H_n(x)$,
$\ps= \sum_{n=0}^\infty c_n e^{-iE_nt}~ <x|n>$, where
$<x|n> = e^{-\half x^2}~ H_n(x)$
$$\eqalign{
W^{(r,s)}=& \sum_{m,n=0}^\infty <n|a^r a^{\dagger s}|m>c^\dagger_m c_n
\cr
= & \sum_{n=0}^\infty <n+s-r|a^r a^{\dagger s}|n>
c^\dagger_{n+s-r}c_n  \cr
}
\eqn\eightseven $$
Note that $\e_{rs}=a^ra^{\dagger s}$ have well defined action on the
Hilbert space of Hermite polynomials.

Now consider the action of $W^{(r,s)}$ on the ground state $|\Omega>$,
which we constructed in section 7. Identical to our discussion in
section 5, we see that the vacuum is an eigenstate of all $W^{(r,r)}$
, with eigenvalues which are moments of the level density $\rho(E)$,
which were originally calculated within this formulation in
Gross-Milkovic [\GROSSMILKOVIC]. For
$r\not=s$, we see that
$$W^{(r,s)}|\Omega>=0, r<s \eqn\eighteight$$
and the state $|r,s>=W^{(r,s)}|\Omega>, r>s$ is an
eigenstate of the hamiltonian
$W^{(1,1)}$ with eigenvalue $e_1+ (r-s)$,
$$W^{(1,1)}|r,s>= e_1+(r-s)|r,s>, \;\; r>s\ge 0 \eqn\eightnine$$

It is important to note that we can indeed generate \underbar{all} the
states of the fermi field theory by acting with the $W^{(r,s)}$'s on the
fermi sea $|\Omega>$ sufficient number of times. The reason is that
all the states of the field theory can be generated from the ground
state by exciting fermions from below the fermi level to above it: if
the initial state of the fermion was $n$ and the final state $m$ then
the elementary excitation is given by the operator
$E_{mn}\equiv c^\dagger_m c_n$. A generic state of the field theory
can therefore be written as a certain number of these $E$-operators
acting on the ground state $|\Omega>$. Now equation \eightseven\
expresses the $W^{(r,s)}$ as a linear combination of the $E_{mn}$`s, it
is easy to show that the $E_{mn}$'s can also be expressed as linear
combinations of the $W^{(r,s)}$'s (i.e. the equation \eightseven\ is
invertible). The easiest way to prove this is to consider the basic
boson operator of the theory: the bilocal operator
$\Phi(x,y) \equiv \Psi^\dagger(x)\Psi(y)$. The $c^\dagger_m c_n$ can
be clearly constructed out of the $\Phi(x,y)$ by taking appropriate
moments, on the other hand $\Phi(x,y)= \sum_{r,s}(-1)^s C(r,s) y^{r-s}
\Psi^\dagger(x) x^s ({\del\over \del x})^r \Psi(x), \quad
C(r,s)= r!/((r-s)!s!)$ which means that
moments of $\Phi(x,y)$ can be constructed out of the generators
$W^{(r,s)}$.

The upshot of the above remarks is that the  Fock space of the
(analytically continued) fermion field theory consitutes one
irreducible representation
of the algebra $\woneplus$. (Clearly any state can be reached from
any other by the operators $E_{mn}$ hence by the operators $W^{(r,s)}$).

The other important remark is that since we have described a complete
list of states in the theory, there are no tachyon states in the
spectrum with coninuously varying energy-momentum. This however is not
in conflict with the fact that the analytically continued theory
meaningfully defines the correlation functions of external tachyons
with continuous momentum.

\chapter{The Classical Limit}

In this section we describe  the semi-classical limit of the fermi
field theory in terms of a fermi fluid in the single-particle  phase
space [\BIPZ,\POLB].
In this limit states of the single-particle theory
are described by small cells of area $2\pi\hbar$ in the 2dim phase
space, and states of the field theory are
described by specifying which of these
single-particle states are occupied. Remembering that each
single-particle  state can accomodate only one fermion, a state in the
field theory corresponds to a region of the 2dim phase space uniformly
filled with a fermi fluid (for simplicity we shall consider a single
connected region). If we describe the curve bounding this region by
the equation $f(x,p)=0$ the function $f$ would then specify the
state of the field theory in the semi-classical limit. \foot{We should
remark that the states which are appropriate for describing the
semi-classical limit are coherent states: in fact in our problem they
are coherent states of the $W$-algebra; the reason is, as we shall see
later we can co-ordinatise the classical phase space in terms of the
classical values of the $W$-generators-- which in particular means
that different non-commuting generators should have sufficiently
well-defined values, which is typical of coherent states of a group.}
Of course, in general the fluid profile $f$
would evolve in time, because each  fermion
would evolve according to the single-particle
hamiltonian, tracing out circles (in the case of the harmonic
oscillator with angular frequency $1$) and unless the profile is a
circle to start with, it will change from $f(p,q)=0$ to $f_t(p,q)=0$.
The dynamics of the classical
field theory consists in solving the motion of the fluid profiles. In
section (9.3) we will explicitly solve and classify all the classical
solutions of the euclidean field theory.

We described how states are represented semiclassically. How about
operators? Since the  fermi field theory is quadratic, it is enough to
consider a generic fermion bilinear $G \equiv \int dx \psd g(x,
-i\del/\del x,t)\ps $. In the semi-classical picture this simply
measures the total amount of $g(x,p,t)$ carried by all fermions in the
fluid. In other words, if we consider a state of the field theory
$|f>$ described by a fluid profile f then we have the Thomas-Fermi
correspondence
$$<G(t)>_f=<f|\int dx g(x,-i\del/\del x,t)|f> \sim {\int\int}_{R_f} dx
dp g(x,p,t) \eqn\ninezero$$
where $R_f$ denotes the region bounded by the fluid profile at time t
$f_t(x,p)=0$.
As we remarked before, the states $|f>$ are coherent states.

We can see how most quantities in the classical field theory are
related to quantities in the single-particle phase space. It is
appropriate therefore to understand the $W$-algebra and in particular
the sub-algebra $SL(2)$ that we found in
the quantum theory in this 2dim classical phase
space.

\section{\underbar{Representation of $SL(2)$ and the $W$-algebra in 2dim
phase space}}

The basic reason why the above algebras appeared in our quantum theory
is that the single-particle classical phase space naturally carries a
representation of $w_\infty$ and even more naturally of $SL(2)$; the
former classically is the set of
area-preserving diffeomorphisms in two dimensions and it is precisely
in two dimensions that the set of area-preserving diffeomorphisms is
also the one which preserves the poisson bracket. Therefore $w_\infty$ is
simply the algebra of all canonical transformations. $SL(2)$ by
definition is the set of real two-by-two matrices of determinant one,
in other words they are linear transformations which preserve area.
Thus $SL(2)$ transformations are canonical transformations, and
therefore the $SL(2)$ subalgebra of $w_\infty$ must be simply those
subset of canonical transformations which act linearly on the
coordinates $(x,p)$! We shall see this now explicitly.

A canonical transformation is generated by taking Poisson bracket with
some function $f(x,p)$:
$$x\to x + \{x,f\},\quad p\to p+ \{p,f\} \eqn\nineone$$
A generic $f$ can be expanded in the basis of the monomials
$f^{(r,s)}= x^r p^s$
The $W^{(r,s)}$ acts on coordinates as
$$W^{(r,s)}: (x,p)\rightarrow  (x+\poi x {f^{(r,s)}}, y+\poi y
{f^{r,s}}) $$
or
$$\delta_{rs}(x,p)= ( s x^r p^{s-1}, - r x^{r-1} p^s) \eqn\ninetwo$$
When the function $f$ is quadratic, the canonical transformation is
linear. In particular, if one takes the basis $f=x^2,-p^2,xp$ of the
quadratics,  then we find from \nineone\ that the corresponding
canonical transformations have the effect of multiplying the column
vector $(x,p)^T$ by the Pauli matrices $\sigma_+,\sigma_-$ and
$\sigma_3$  respectively. This explains very naturally why $SL(2)$
appears in $w_\infty$, and specifically as $W^{(2,0)}, W^{(1,1)}$ and
$W^{(0,2)}$.

Commuting two $W$-flows corresponds to taking Poisson brackets of the
corresponding generating functions. The $f^{(r,s)}$'s form a
closed Poisson bracket algebra:
$$ \poi {f^{r,s}} {f^{p,q}}= (rq- sp) f^{r+p-1, s+q-1}
\eqn\ninethree$$

It's interesting to see
how the special $SL(2)$
sub-algebra acts on the rest of the generators. \ninethree\ gives
$$\eqalign {
\poi {x^2} {x^r p^s}=& 2 s x^{r+1} p^{s-1} \cr
\poi {p^2} {x^r p^s}=& -2 r x^{r-1}p^{s+1} \cr
\poi {xp}  {x^r p^s}=& -(r-s) x^r p^s           \cr
} \eqn\ninefour $$

The first observation is that, the degree of a polynomial (assumed
homogeneous in $x,p$) is preserved under Poisson bracket with any of
the $SL(2)$-generators. Thus, for instance the monomials $\{ p^n, x
p^{n-1}, \cdots, x^{n-1} p, x^n\}$ are transformed into one another by
the $SL(2)$ action, where all of these have degree $n$. Clearly there
are $2n+1$ such monomials of degree $n$, thus the dimension of the
$SL(2)$ representation is $2n+1$. This tells us that the $j$-value
(Casimir= $j(j+1)$) for the representation is $j=n$, simply the degree
of the Polynomial (it is easy to see the above
representation is irreducible, for
instance the action of $x^2$ takes one down from any basis element
to the next).  For later use, we remark that the degree of a plynomial
in $x,p$ is represented by the operator
$$J=x\del/\del x+ p\del/ \del p \eqn\ninefive $$
Thus $J$ measures the $j$-spin of a representation.

Thus we conclude that the entire $W$-algebra splits into irreps of the
$SL(2)$ with $r+s$ being the spin  $j$ and $r-s$ being the
$J_3$-value (remember in \ninefour\ the operator $xp$ plays the role
of $J_3$ and the other two are the rasing and lowering operators
$J_\pm$).

In the next section, we shall consider the $W^{(r,s)}$ generators in
the semi-classical field theory.
{}From the next section, in order to
correspond exactly to the earlier discussion in the quantum theory
operators we shall redefine our $W^{(r,s)}$ in terms of functions
$$g^{(r,s)}= (p-ix)^r (p+ix)^s\eqn\ninefive$$.

\section{\underbar{Correspondence with vertex operators of Liouville theory}}

Let us examine the classical limit of the generator $W^{(r,s)}$. By
the Thomas-Fermi correspondence \ninezero\ we have

 $$<W^{(r,s)}(t)>_f \equiv w^{(r,s)}(t,f)
=\int\int_{R_f}  (p-ix)^r (p+ix)^s \eqn\ninesix$$

where we have  put $g= g^{(r,s)}$ (equation \ninefive) in \ninezero.

Introducing the polar co-ordinates $p=R\cos\theta, x=R\sin\theta$
\ninesix\ becomes
$$ w^{(r,s)}(t,f)= \int\int_{R_f}
d\theta dR R R^{r+s} e^{i(r-s)(t- \theta)}
 \eqn\nineseven$$

Parametrising the radius $R= \exp (-\tau), -\infty <\tau< \infty$,
\nineseven\ becomes
$$ w^{(r,s)}(t,f)= \int \int_{R_f} d\theta d\tau V^{(r,s)}(\tau, t-
\theta ) \eqn\nineeight $$
where
$$V^{(r,s)}(\tau,t)= e^{-2 \tau- (r+s)\tau + i(r-s)t} \eqn\ninenine$$
It is intriguing to note that
this is identical to the exponential part of the $r,s$ vertex
operators in the $c=1$ Liouville theory for one of the liouville dressings,
where the time $t$ is taken
to be Euclidean. Using \ninefive\ we see that the operator $J$
which measures the total
$j$-spin is $-\del/\del\tau$.  Since the $J$-spin coincides with
Liouville momentum the identification of $\tau$ as the Liouville
direction is natural.
(To get the more  standard expression we need to put
$ \tau= \phi/ \sqrt{2}$ and rescale $t \to t/ \sqrt{2}$).
Importantly
we are finding here vertex operators with only quantised energies,
implying in partiulcar the absence of the tachyon vertex operators
with continuously varying energy and momentum. The operators $r,0$ and
$0,s$ corresponds to the tachyon (energy$^2$ + momentum$^2$=0) but
only at integral values of the energy.

\section{\underbar{The classical solutions
and the classical phase space of the field theory}}

We have argued in the beginning of this section that the semiclassical
dynamics of the fermi field theory is determined by the motion of a
fluid  of uniform density in the single-particle phase space
parametrised, for instance, by the equation $f=0$ for its boundary
(which we shall call its ``profile''). This motion in turn is
determined by how each fermi particle moves in the single-particle
phase space under hamiltonian evolution.
Let us ask the question how to solve for the motion of
all possible profiles $f$
as a function of time. In other words, given a fluid boundary
$f(x,p)=0$ at
time $t=0$, what is the function $f_t$ such that $f_t(x,p)=0$
describes the fluid boundary at time $t$.

The answer  is suprisingly simple:
$$ f_t(x,p)= f(x \cos t+ p \sin t, -x \sin t + p \cos t) \eqno
\eqname\clone$$
To prove it, one just needs to remember that
\underbar{all} the fermions move in circles in the phase space
under the single-particle hamiltonian at an identical angular speed
(which we have taken to be 1). Hence if you look from a rotating frame
of angular speed one, all the particles would look static and so would
the fluid boundary. In other words, the fluid moves like a rigid body
in a two-dimensional plane rotating round the origin.

Let us point out one immediate consequence of the formula \clone.

\underbar{All classical solutions of the theory are periodic}.

Proof: $f_t(x,p)= f_{t+ 2\pi}(x,p)$ irrespective of the initial
function $f$.  Q.E.D

Of course, this is intimately connected with the fact that we are
talking here about the Euclidean field theory where the
single-particle orbits have become periodic in phase space (instead
of parabolic).

\underbar{The space of classical solutions as the classical phase space}:

Equation \clone\ gives in principle all the classical solutions of the
system. Since the space of classical solutions is the most natural
definition of the classical phase space of our field theory (we shall
call this phase space $M$) let us
try to understand it better. We can parametrise the space of functions
$f(x,p)$ as
$$f(x,p)= \sum_{mn} a_{mn}x^m p^n= \sum_{mn}\alpha_{mn}(p+ix)^m
(p-ix)^n, \quad \alpha_{nm}= (\alpha_{mn})^* \eqn\cltwo$$
We can think of $\alpha_{mn}$ (or equivalently $a_{mn}$) as parametrising
\foot{
a few remarks are in order:
in general the fermi fluid may have several
disconnected regions; this would seem to require several functions
$f_1(x,p), f_2(x,p),\cdots$ to specify the boundaries of the different
disconnected regions; however interestingly the union of the contours
of $f_i(x,p)=0, i=1,2,\cdots$ is the same as the contour  of
$f(x,p)=0$ where $f(x,p)= \prod_i f_i(x,p)$. In other words we are not
losing on generality when we describe the space of fermi fluids by a
single function $f(x,p)$. The second point is, the $\alpha_{mn}$'s
are actually constrained, by the requirement that the fermi fluid
should always occupy the same volume. For our purposes, this would
not be too important, because we will consider flows in this
space that automatically preserve this constraint. Lastly, the
$\alpha_{mn}$'s contain some redundant information since the function
$f$ can be deformed preserving its zero contour; so different
$\alpha_{mn}$'s may actually refer to the same point. A better
coordinatisation is provided by the functions  $w^{rs}$ on $M$ which
we will describe shortly.}
$M$.

We would like to make a comment here about bosonisation. As Polchinski
has shown, in case the fluid profile is quadratic in $p$, i.e.
$f(x,p)= (p-p_+(x))(p-p_-(x))$ the upper and lower boundaries
$p_\pm(x)$ behave as canonically conjugate variables of a classical
bosonic field theory. The quadratic restriction on the profile,
however, is rather unnatural, as is clear from the fact that such a
restriction is not even preserved in time. For instance if we start
with a profile $f=p^2+ x^4-1=0$ at time $t=0$ after a quarter period
$t=\pi/2$ the profile looks like $f(t=\pi/2)=x^2+ p^4-1=0$ which is no
longer quadratic in $p$. We would
therefore like to propose the entire space of fluid profiles (all the
$\alpha_{mn}$`s)  as the
correct bosonic variables  for our problem. In retrospect, this is a
rather natural route to bosonisation for a fermi system: (a) find the
space of classical solutions, (b) identify it as a classical phase
space
and (c)try to quantise the variables of the system by replacing
Poisson
brackets with commutators (the observables are naturally bosonic in
this procedure).

So we have the  $\alpha_{mn}$'s as the phase space variables of a
classical bosonic system. By the Thomas-Fermi correspondence
$\ninezero$, we know the  classical hamiltonian as a function of the
$\alpha_{mn}$:
$$H(\{\alpha_{mn}\})=\int\int_{R_f} dx dp \; (p^2+ x^2)/2
\eqn\clthree$$
The functional dependence on the $\alpha_{mn}$ comes from the
integration region $R_f$ specified by the boundary $f=\sum
\alpha_{mn}z^m
\zb^n=0$, where $z = p + ix,~\zb = p-ix$.
This is a rather implicit-looking function; but we'll find
that we dont need to know it exactly.

If we tried to quantise the space $M$ of the $\alpha_{mn}$'s we would
like to know, for instance, the classical orbits in the space $M$, and
how to compute Possion brackets of two functions $F(\{\alpha_{mn}\})$
and $G(\{\alpha_{mn}\})$.

{\sl Classical orbits:}

We simply transcribe the evolution $f\to f_t$, described in \clone, as
$\alpha_{mn}\to \alpha_{mn}(t)$ ($\alpha_{mn}(t)$ defined by $f_t(x,p)
= \sum_{mn} \alpha_{mn}(t) z^m \zb^n$, thus $\alpha_{mn}=\alpha_{mn}
(0)$).  The answer is:
$$\alpha_{mn}(t)= \alpha_{mn}(0) \exp(i (m-n)t) \eqn\ninefour$$

This again is the list of all classical orbits in the Euclidian phase
space. Note that all classical solutions are periodic and have integer
frequencies.

{\sl Poisson brackets:}

Equation \ninefour\ tells us about the Poisson bracket of the
$\alpha_{mn}$ and $H(\{\alpha_{pq}\})$:
$$\poi{\alpha_{mn}} {H}= {\del\over \del t}|_{t=0}\alpha_{mn}(t)=
i(m-n)\alpha_{mn} \eqn\ninefive$$

Note that we have computed this bracket without either knowing the
explicit functional form of $H(\{\alpha_{mn}\})$ or $\poi
{\alpha_{mn}} {\alpha_{pq}}$. The fermi fluid picture
told us how to find the
flow of the point $\alpha_{mn}$ under the hamiltonian, and that gave
us the Poisson bracket directly.

If we can calculate the  Poisson brackets of two arbitrary functions
on  $M$ by this method,
we wont need to know the Poisson bracket $\poi
{\alpha_{mn}} {\alpha_{pq}}$.

By the Thomas-Fermi correspondence \ninesix, each of the
$w^{rs}$'s correspond to a specific function of the
$\alpha_{mn}$'s. In the quantum theory, we saw that all observables
can be built from the $W^{(r,s)}$'s and their products (since the basic
fermion bilinear $c^\dagger_m c_n$ could be written in terms of the
$W^{(r,s)}$'s). In the classical theory, that would mean that all
functions on $M$ can be expressed as functions of the $w^{rs}$. Thus
it is enough to compute the Poisson bracket $\poi{w^{rs}}{w^{pq}}$.

To do this, think of the Poisson bracket as the result of commuting
two independent flows. Each of these flows is induced on $M$ from a
corresponding flow in the single-particle phase space $Q$.
To be precise the evolution in $M$ under
$w^{rs}$ is computed by how the fluid profile evolves when the single
fermions are evolved by a  ``hamiltonian''
$(p-ix)^r (p+ix)^s$. (One can show that this implies a Poisson bracket
$\poi{\alpha_{mn}}{w^{rs}}= ((m+1)s-(n+1)r)\alpha_{m+1-r,n+1-s}$).
{}From this map between flows, it is easy to see
that the Poisson bracket algebras are isomorphic.

We therefore have the result:
$$\poi {w^{rs}}{w^{pq}}= (rq-sp) w^{r+p-1,s+q-1} \eqn\ninefive $$
Remembering that $H=w^{11}$ and that all other functions on $M$ are
functions of the basic observables $w^{rs}$, \ninefive\ specifies the
classical physics completely. In principle we now have all the
ingredients to quantize the  theory by doing functional
integral over the classical phase space.

Let us explain the last comment in a little more detail.
Consider the
simpler example of a finite ($n$) dimensional phase space $M_n$, with
generalised coordinates $\xi^i$. Suppose one has $n$ independent
functions $f^i$ on $M$ whose Poisson brackets  are known, namely
$$\poi{f^i}{f^k}= \Omega^{ik}(\{f^j\})$$
Also suppose the hamiltonian $h(\{\xi^i\})$ depends on the $\xi^i$
through the functions $f^i$: $h= h(f^i(\xi^k))$. To write down a
quantum theory first one needs to construct the symplectic form, which
can be easily shown to be $\Omega= \Omega_{ik} df^i\wedge df^k$ where
$\Omega_{ik} $ and $\Omega^{ik}$ are inverse matrices. Since the
symplectic form is closed (can be derived from the Jacobi identity for
Poisson brackets) locally one can find a one-form $\theta= \theta_i
df^i$ such that $d\theta= \Omega$. The phase space functional integral
can be written by defining the lagrangian $\theta_i {\dot {f^i}}-
h(\{f^j\}) $ and integrating with respect to a measure consistent with
the volume form $\Omega= \Omega^{n/2}$.

In our problem, the $w^{rs}$ are not independent functions, because the
phase space $M$ is not the group $G$ (of
area-preserving diffeomorphisms) itself, but rather a coset $G/H$. Thus,
for instance
if one starts with a fermi fluid which corresponds to the filled fermi
sea with the circular symmetry, then the action of all the generators
$w^{rr}$ keep it invariant. Therefore there is a non-trivial isotropy
subgroup $H$ formed by the diagonal generators. A particular choice of
independent functions could be $w^{rs}, r \not= s$. (For example on
$S^2= SO(3)/SO(2)$ the functions  $J^\pm, J^3$ are not all independent,
as evident from the particular representation $J^+= z^2, J^-= \zb^2, J^3
= z\zb$ which says $J^3= \sqrt{ J^+ J^-}$).
Thus \ninefive\ gives us the matrix $\Omega^{ik}$ mentioned in the last
paragraph which gets written as $ \Omega^{rs, pq} (\{
w^{kl},k\not=l\}) = (rq - sp) w^{r+p-1, s+q-1}$.

In the space of the non-diagonal $w^{rs}$'s this matrix is invertible.
Let's call the inverse matrix (counterpart of $\Omega_{ik}$)
$ \Omega_{rs, pq}(\{w_{kl}, k\not= l\})$ and the one-form $\theta_i$ as
$\theta_{rs}(\{w_{kl}, k\not= l)$.
The phase space action  ($p \dot q- h(p,q)$) is therefore given by
$$L= \theta_{rs} (\{ w^{kl} \}){d w^{rs}\over dt} -w^{11}(\{ w^{kl}\})
\eqn\ninefivea$$
where by the set $\{w^{kl}\}$ we mean again the off-diagonal $w$'s and the
function $w^{11}(\{w^{kl}\})$ is determined by its poisson bracket
with the $w^{kl}$.
\footnote*{
As an example of this procedure, we can think of $S^2$ as coordinatised
by $J^+, J^-$. The matrix $\Omega^{ik}$ has non-zero elements $\Omega^{12}
= - \Omega^{21}= J^3= \sqrt{ J^+ J^-}$ whose inverse is given by
$\Omega_{12}= - \Omega_{21}= 1/ J^3$. This gives $\theta= \sqrt{J^+} d(
\sqrt {J^-})$ implying $L= \sqrt{ J^+} \dot {J^-} - \sqrt{ J^+ J^-}$ for the
problem when the hamiltonian is $J^3$. }

It is significant to note that our problem with a symplectic manifold
$W^{(+)}_{1+\infty} / H$ can be formulated as problem involving motion on
$W^{(+)}_{1+\infty}$. This is along the same lines as the monopole problem,
which has a symplectic manifold $SU(2) / U(1)$, and can be formulated as
motion on the group $SU(2)$ using a construction that mimics the
Wess-Zumino term in higher dimensions [\BALA]. Finally we remark that it
would be of interest to study the above action for an arbitrary single
particle Hamiltonian because these correspond to different background
gauge fields. An important question to answer is which single partical
Hamiltonian corresponds to the black-hole background.

\chapter{General background for the   $c=1$ matrix model:}

Let us begin the discussion for any arbitrary background gauge field
$\bA(t)$. As we have seen in section 5, the generators \newfivefour\
of the special gauge transformations that leave the background
invariant are constants of motion. Now let us consider deforming the
(analytically continued) model by these constants of motion:
$$S_\e =  \int dt <\psi(t)| i\del_t+ \bA(t)|\psi(t)>
        +\int dt <\psi(t)| \e(t)|\psi(t)> \eqn\tenone $$
where $\e(t)$ satisfies
$$i\del_t\e+ {\rm ad}\; \bA(t)\; \e=0 $$
Deforming the model at the background $\bA(t)$
according to \tenone\ corresponds to a shift of the background $\bA
\to  \bA+\e$. It is easy to see that $\bA(t)+\e$ is not a gauge
transform of $\bA(t)$. In the case of the specific background $\bA=
(p^2+ x^2)/2$, the time dependence of $\e(t)$ is explicitly known and
hence \tenone\ becomes
$$ S_\e= \int dt dx \psd (i\del_t+ {p^2 + x^2\over 2})\ps + \sum_{r,s}
g_{rs} W^{(r,s)} \eqn\tentwo $$
where
$$W^{(r,s)}= \int dt dx \exp i(r-s)t\quad \psd a^r a^{\dagger s}\ps
\eqn\tenthree $$
and we have defined the couplings $g_{rs}$ by
$$\e(t)= \sum_{r,s} g_{rs} \exp i(r-s)t a^r a^{\dagger s} $$
Note that these couplings (which correspond to higher spin fields in
space time) are conjugate to the generators $W^{(r,s)}$ of the
infinite dimensional algebra $\woneplus$.

As yet our formulation of the problem has been such that a background
gauge field $\bA(t)$ is given to us. The main question is what
principle determines these backgrounds.

One possibility is to note that our specific background $\bA= (p^2 +
x^2)/2 $ led to a specific representation of the algebra $\woneplus$
whose explicit construction was given in section 8, using the
eigenfunctions and eigenvalues of the operator $\bA= (p^2+ x^2)/2$.
This points to the fact that for each background $\bA(t)$ there will
be an explicit representation of $\woneplus$. Hence if by some means
we can classify and understand all representations of $\woneplus$,
then we will have classified and understood all the background gauge
fields $\bA(t)$. We believe that this is an atractive scenario. It
would be most interesting to discover a background or representation
that for instance describes the black hole solution of two-dimensional
string theory.

One way of approaching the problem of representation is by appealing
to the classical phase space  discussed in section 9.

\chapter{ Connection with higher  dimensional field theories}

It is clear from the above discussions that our gauge-invariant
lagrangian $L(\psi,\psi^\dagger, A)$ is symmetric under the group
$F(R, G)$ where $G$ is the group of unitary (in the classical limit,
canonical) transformations on the single-particle hilbert space
and $F(R,G)$ denotes
all maps from the real line (time) to $G$. The periodic maps are
a particular sunbgroup of this, which is of course the loop group $LG$.
The loop group $LG$ is particularly relevant for the background of $A$
which corresponds to the harmonic oscillator potential, because in
that case the classical solutions are periodic in time, and one
classical solution is transformed to another by the action of $LG$
rather than the full group $F(R,G)$.

Now interestingly, the loop group $LG$ is also the group of
transformations from one classical solution to another for $N=2$ strings
living in a 4-dimensional space of (2,2) signature[\VAFA]. In other
words, the classical phase space of $N=2$ string field theory carries a
natural representation of $LG$; indeed each classical solution can be
mapped onto a specific element of $LG$ which can be denoted by a
function $f(t,x,p)$ where $f$ is an arbitrary function of $t$ and the
two-dimensional plane $x,p$ (at each $t$, $f$ can be used to generate a
canonical transformation on the two-dimensional plane).  Each solution
of the $N=2$ SFT can also be shown to be correspond to a solution of
self-dual Einstein gravity (with (2,2) signature).

Thus, we see that in terms of classical phase spaces,  our fermion-gauge
field system is closely related to $N=2$ string field theory and also
self-dual Einstein gravity in four dimensions.

\bigskip
\noindent{\bf  Acknowledgements:}
We would like to thank L. Chandran, K.S. Narain, S. Rajeev and
A.M. Sengupta for useful discussions. We would also like to thank
S. Rajeev for bringing reference [\RAJV] to our attention.
The motivation to discuss our results in the limiting case of the
classical fermi fluid arose from a discussion with E.Witten. We would
like to thank him for that. G.M. and S.R.W. are also indebted to him
for numerous conversations on various aspects of the subject of this
paper.
S.W. would like to thank J.Frohlich for hospitality at the ETH
where part of the work was done. The work of G.M.
and S.W. was supported in part by a Department of Energy grant
DE-FG02-90ER40542.

\noindent Note Added: Although the operators $W^{(r,s)}$ are not defined in
the Minkowski theory, one could introduce a generalization of the loop
operator which provides a "regularized" definition of the $W^{(r,s)}$ much
in the same way in which the loop operator provides a "regularized"
definition for the moments of the density operator [\MORE]. This
generalization of the loop operator is
$$ W(p,q,t)={1\over2}\inx~e^{ipx}\psi^{+}(x+{1\over2}q,t)
\psi(x-{1\over2}q,t)$$
Formally the $W^{(r,s)}$ can be obtained from this by taking appropriate
number of derivatives with respect to $p$ and $q$ at $p=q=0$. These
operators satisfy the algebra
$$[ W(p,q,t), W(p',q',t) ]=i~ sin{1\over2}(pq'-p'q)~
W(p+p',q+q',t)$$
One can derive a closed set of Ward identities for these operators which
are first order partial differential equations in $p,q,t$ and relate the
$n+1$ point function to the $n$ point function. Since the one point
function can be evaluated exactly using the parabolic cylinder functions,
in principle the higher point functions can be obtained by solving the
differential equations mentioned above. We have solved for the two point
function of the $W(p,q,t)$'s this way and obtained the two point functions
of the $W^{(r,s)}$'s. These show poles at precisely the expected energies.

One could also introduce a generating functional for the $n$ point
functions of the $W(p,q,t)$'s by introducing sources in the fermionic
action. One can show that this generating functional satisfies a Ward
identity coming from the algebra of the $W(p,q,t)$'s. This Ward identity
can be solved for the generating functional perturbatively in the sources.
The Legendre transform of the generating functional is presumably the
quantum effective action for the discrete states. The details of this will
appear elsewhere.

\bigskip
\noindent
Figure captions:

\noindent Fig 1a. Fermi surface: $x^2- p^2=\mu$.

\noindent Fig 1b. Fermi surface: $x^2+ p^2=\mu$.

\refout
\end